\documentclass{article}

\PassOptionsToPackage{numbers, compress}{natbib}
\usepackage[final]{nips_2017}

\usepackage{graphicx}
\usepackage[T1]{fontenc}    % use 8-bit T1 fonts
\usepackage[hidelinks]{hyperref}       % hyperlinks
\usepackage{url}            % simple URL typesetting
\usepackage{booktabs}       % professional-quality tables
\usepackage{amsfonts}       % blackboard math symbols
\usepackage{nicefrac}       % compact symbols for 1/2, etc.
\usepackage{microtype}      % microtypography
\usepackage{listings}
\usepackage{xcolor}
\usepackage{courier}
\title{An introduction to approximate computing\thanks{An earlier version of this article first appeared as a blog post: \url{https://blog.formallyapplied.com/2017/11/approx-intro/}}}

\author{
  M. Ammar Ben Khadra \\
Department of Electrical \& Computer Engineering\\
Univesity of Kaiserslautern, Germany\\
\texttt{khadra@eit.uni-kl.de}\\ 
}

\begin{document}

\maketitle

\begin{abstract}
Approximate computing is a research area where we investigate a wide spectrum of techniques
to trade off computation accuracy for better performance or energy consumption.
In this work, we provide a general introduction to approximate computing.
Also, we propose a taxonomy to make it easier to discuss 
the merits of different approximation techniques. 
Our taxonomy emphasizes the expected cost of tackling approximate computing across the entire system stack. 
We conclude by discussing the unique opportunities as well as challenges of  nondeterministic approximate computing.

\end{abstract}

\section{Introduction}
Approximate Computing (AC) is a wide spectrum of techniques that relax the accuracy of computation in order to improve performance, energy, and/or another metric of interest. AC exploits the fact that several important applications, like machine learning and multimedia processing, do not necessarily need to produce precise results to be useful.
For instance, we can use a lower resolution image encoder in applications where high-quality images are not necessary. In a large-scale data center, such approximation may lead to vast  savings in the required amount of processing, storage, and communication bandwidth.

Research interest in approximate computing has been growing in recent years motivated by its potential in reducing power consumption. 
Additionally, research has gained even more traction due to the rising concerns over microchip reliability as CMOS technology moves to 7 nm processes and beyond. 
We refer here to two recent surveys \cite{Mittal2016,Xu2016} for a comprehensive treatment. 
In this article, we'll try to provide an introduction to this research area. Key concepts will be overviewed based on a proposed taxonomy. We further elaborate on nondeterministic AC, an AC category with unique opportunities as well as challenges.
Graduate students will, hopefully, find this introduction useful to systematize current research and identify interesting problems.

Note that some of the ideas discussed here are based on an earlier work \cite{Benkhadra2016b}. In our extended abstract, we highlighted some AC challenges and opportunities in general. There, we argued that nondeterministic AC faces a fundamental control-flow wall which is bad news! The good news, however, is that there are still large opportunities in deterministic AC to keep practitioners busy for the foreseeable future.
We shall elaborate on those issues and more in the following. Our discussion is structured as follows. First, we motivate the need for research in AC. Next, we attempt to define the scope and main goals of AC. This is particularly important due to the inconsistencies that can be found in the literature. Later, we propose a general taxonomy which makes it easier to discuss the merits of different AC techniques. 
Our taxonomy emphasizes the \textit{cost} expected for tackling AC across the entire system stack.
Finally, we focus on nondeterministic AC due to its unique opportunities and challenges.

%%%%%%%%%%%%%%%%%%%%%%%%%%%%%%%%%%%%%%%%%%%%%%
\section{Research motivation}  
%%%%%%%%%%%%%%%%%%%%%%%%%%%%%%%%%%%%%%%%%%%%%%

So let's start by trying to answer the \textit{why} and \textit{what} questions for AC. In this regards, our concrete questions will be:

\begin{itemize}
	\item What motivates the recent academic interest in AC? In other words why do we need to care about AC more than before?

\item What approximate computing really is in the first place? Note that AC is used in practice for decades already, so what makes recent academic proposals different?

\end{itemize}

Research in AC can be motivated by two key concerns, namely, power and reliability. 
We shall address both in the following.
Nowadays, the majority of our computations are being done either on mobile devices or in large data centers (think of cloud computing). Both platforms are sensitive to power consumption. That is, it would be nice if we can extend the operation time of smartphones and other battery powered devices before the next recharge.
Also, and perhaps more importantly, the energy costs incurred on data centers need to be reduced as much as possible. Note that power is one of (if not the) major operational cost of running a data center. To this end, algorithmic optimizations, dynamic run-time adaptation, and various types of hardware accelerators are used in practice.
Vector processors, FPGAs, GPUs, and even ASICs (like Google's recent TPU\footnote{Tensor Processing Unit: \url{https://en.wikipedia.org/wiki/Tensor_processing_unit}}) are all being deployed in an orchestrated effort to optimize performance and reduce power consumption.

In this regard, the question that the AC community is trying to address is; can we exploit the inherit approximate results of some application to gain even more power savings?

We move now to reliability concerns to further motivate AC. The semiconductor industry is aggressively improving it's production processes to keep pace with the venerable Moore's law (or maybe a bit slower version of it in recent years). Microchips produced with 10 nanometer processes are already shipping to production. Moving toward 7 nanometers and beyond is expected within the next 5 years according to the ITRS roadmap\footnote{International Technology Roadmap for Semiconductors: \url{http://www.itrs2.net/}}.
Such nanometer regimes are expected to cause a two-fold problem. First, transistors can be more susceptible to both temporary and permanent faults. For example, cosmic radiation can more easily cause a glitch in data stored in a Flip-Flop. Consequently, more investment might be necessary in hardware fault avoidance, detection, and recovery.
Second, feature variability between microchips, or even across the same microchip, can also increase. This means that design margins need to be more pessimistic to account for such large variability. That is, manufacturers need to allow for a wider margin for the supply voltage, frequency, and other operational parameters in order to keep the chip reliable while maintaining an economical yield.

In response to these reliability challenges, the AC community is investigating the potential of using narrower margins to operate microchips. However, hardware faults might occasionally propagate to software in this case. Therefore, the research goal is to ensure that such faults do not cause program outputs to diverge too much from the ideal outputs.
Such schemes can allow chip manufacturers to relax their investments in maintaining hardware reliability. More performance and power saving opportunities can be harnessed by moving to best-effort hardware instead \cite{Chakradhar2010}.

%%%%%%%%%%%%%%%%%%%%%%%%%%%%%%%%%%%%%%%%%%%%%%
\section{Definition and scope}
%%%%%%%%%%%%%%%%%%%%%%%%%%%%%%%%%%%%%%%%%%%%%%

So far, we motivated the need for approximate computing. We move now to our second question; What is approximate computing in the first place?

Let's come back to our image encoding example. We know that applying any lossy compression algorithm (JPEG for example) to a raw image will result in an approximate image. Such compression often comes - by design - with little human perceptible loss in image quality. Also, image encoders usually have tunable algorithmic knobs, like compression level, to trade off image size with its quality.
Therefore, do not we already know how to do AC on images? Actually, instances of AC are by no means limited to image processing. AC is visible in many domains from wireless communication to control systems and beyond. In fact, one can argue that every computing system ever built did require balancing trade-offs between cost and quality of results. This is what engineering is about after all.

Indeed, approximate computing is already a stable tool in the engineering toolbox. However, it is usually applied \textit{manually} leveraging domain knowledge and past experience. The main goal of recent AC research is to introduce \textit{automation} to the approximation process. That is, the research goal is to (semi-) automatically derive/synthesize more efficient computing systems which produce approximate results that are \textit{acceptable}.

\begin{figure}[h!]
	\centering
	\includegraphics[width=0.6\linewidth]{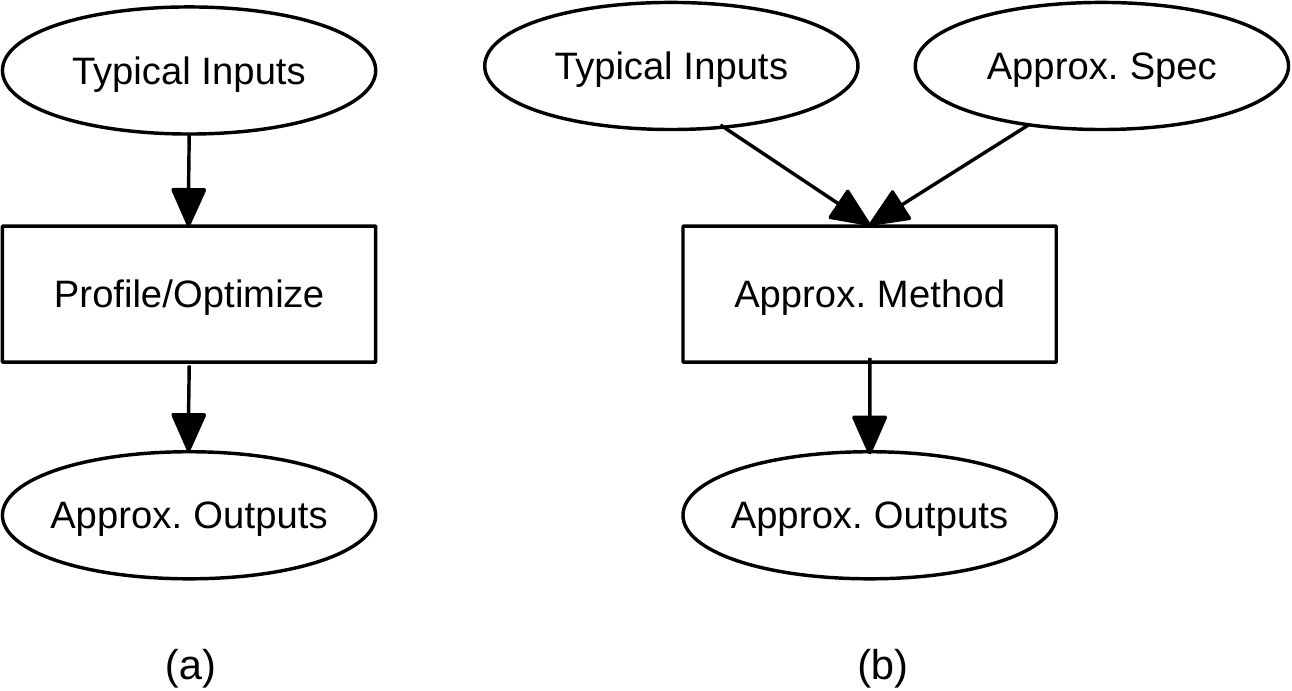}
	\caption{AC can be applied (a) manually in the usual profile-optimize cycle, or (b) automatically via an approximation method which may require providing an error quality specification.}
	\label{fig:intro}
\end{figure}

We elaborate on this point based on Figure \ref{fig:intro}. Consider for example that you have been given a computing system with a well-specified functionality. Such system can consist of software, hardware, or a combination of both. Now, your task is to optimize this system to improve its performance. How would you typically go about this task?
Ideally, you start by collecting typical inputs which represent what you expect the system to handle in the real world. Then, based on those inputs, you attempt to profile the system to identify \textit{hot regions} where the system spends most of its time. After that, the serious optimization work begins which might involve several system layers.

For example, in a software program, you will often need to modify the algorithms and data structures used. You might also go all the way down to the nitty-gritty details of improving cache alignment or ``stealing'' unused bits in some data structures for other purposes.

This profile-optimize cycle continues until you either meet your performance target or you think that you have reached the point of diminishing returns. This optimization process is generally applicable to any computing system. However, you can go a bit further in your optimization given a system that can tolerate controllable deviations from its original outputs.

Basically, AC is about this last mile in optimization. The research goal is to investigate \textit{automatic}, \textit{principled}, and ideally \textit{generic} techniques to gain more efficiency by relaxing the exactness of outputs. The need for automation is obvious since manual approximation techniques can simply be regarded as ``business as usual'', i.e., without clear improvement over the state of practice.
Also, AC needs to provide guarantees, in a principled way, that expected output errors will remain ``acceptable'' in the field. That is, computing systems already struggle with implementation bugs. Therefore, it is difficult to adopt an AC technique that can introduce more bugs in the form of arbitrary outputs.

We come to the third criterion which is generality. We consider a technique to be generic if it is applicable to a wide spectrum of domains of interest to AC. For example, loop invariant code motion is a generic compiler optimization applicable to virtually any program from scientific simulations to high-level synthesis.
Generality, however, is more challenging to achieve in AC compared to the ``safe'' optimizations used in compilers. Basically, a local AC optimization might introduce errors that are difficult to reason about when combined with other local AC optimizations. Note that the combined error observed on the global outputs might be composed of several local errors.

Actually, we would argue that it is not feasible to target, to a satisfactory level, all three criteria. In other words, better automation and principled guarantees require compromising on generality. This can be achieved by embedding domain-specific knowledge in the AC technique. This seems to be a reasonable thing to do given the diversity of domains where AC is applicable.
A prime advantage of compromising on generality is that end users won't need to explicitly provide error quality specification, see Figure \ref{fig:intro}. Metrics of acceptable errors will be based on the specific domain. For instance, generalization error in machine learning and PSNR in image compression.

%%%%%%%%%%%%%%%%%%%%%%%%%%%%%%%%%%%%%%%%%%%%%%
\section{A taxonomy of approximate computing}
\begin{figure}[h!]
	\centering
	\includegraphics[width=0.5\linewidth]{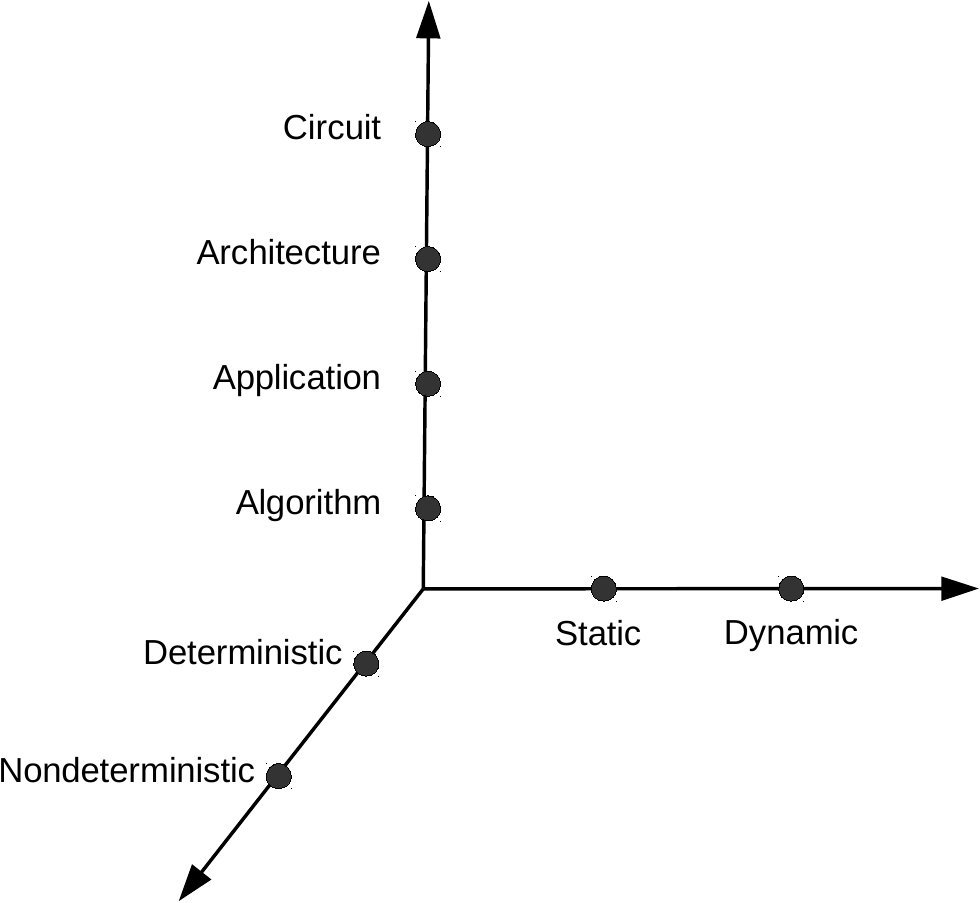}
	\caption{Proposed AC taxonomy. Expected cost of targeting a design point increases as we move away from the center.}
	\label{fig:taxonomy}
\end{figure}

%%%%%%%%%%%%%%%%%%%%%%%%%%%%%%%%%%%%%%%%%%%%%%
The literature on approximate computing is large and growing. Also, it covers the entire system stack from high-level algorithms down to individual hardware circuits. It is difficult to make sense of all of these developments without introducing some sort of structure. In this section, we attempt such structuring based on the taxonomy depicted in Figure \ref{fig:taxonomy}. Also, selected pointers to the literature will be provided.
Basically, our taxonomy is based on the hypothesis that we can map any individual AC technique to a point in a three-dimensional space. The considered axes represent the approximation level, required run-time support, and behavior determinism respectively. Note that there are papers on AC that combine several techniques in one proposal.

The reader might be wondering why hardware circuits have been placed higher up while the algorithm level is at the bottom? The reason is simply the expected cost of targeting such a design point. In other words, we expect the implementation cost to increase as one explores design points further away from the center.
However, cost should always be considered in combination with the value gained. A system that involves dynamic run-time adaption, e.g., for error quality monitoring, is more complex, and thus more costly, to build and maintain compared to a static system. However, the former might provide sufficient benefits to amortize the higher cost if designed with care.

Now, we move to the determinism axis. Our classification is based on the usual determinism property. That is, an algorithm that returns the same output repeatedly given the same input is deterministic. Nondeterministic algorithms do not exhibit such output repeatability.
We further classify nondeterministic algorithms to \textit{partially} and \textit{fully} nondeterministic. These categories are based on the feasibility of accounting for
the sources of nondeterminism a priori. Nondeterministic AC will be discussed in more detail in the next section.

The axis of approximation level in Figure \ref{fig:taxonomy} has been (roughly) divided into 4 categories. At algorithm level, a given algorithm is kept intact. To implement approximation, one has to manipulate either the inputs or algorithm configurations (knobs or hyperparameters). An example of the former can be found in ApproxHadoop \cite{Goiri2015} where the authors utilized a statistical input sampling scheme in order to derive approximate results.
In comparison, one can leave the inputs and modify the hyperparameters instead.
Noteworthy in this category is Capri \cite{Sui2016}. There, the authors formulate knob tuning as a constrained optimization problem. To solve this problem, their proposed system learns cost and error models using bayesian networks.
Note that approximation via hyperparameter optimization is a well-established research theme in machine learning. It is known there as ``learning to learn'' or ``meta-learning''. We won't elaborate on this here and refer the reader instead to \cite{Andrychowicz2016} simply because the paper title seems ``meta'' enough.

Let's move now to application-level approximation where we need to look \textit{inside} the original algorithm. Compare this to the previous algorithm-level approximation where the original algorithm was kept intact as a black-box. Application-level approximation can be obtained by (1) modifying the original algorithm according to some predefined rules or (2)  program synthesis which enables exploring a larger search space.  A good example of (1) is loop perforation  \cite{Sidiroglou-Douskos2011}. There, the authors identified certain loop patterns and proposed techniques to automatically skip loop iterations. That is, instead of running a loop $ N $ times, one can run it  $ M $ times where $ M < N $. As for category (2) we refer to program synthesis via Monte-Carlo sampling as proposed in \cite{Schkufza2013}. 
The authors provide an automatic and principled technique for approximating floating-point
functions which is rather interesting. However, the scalability of their technique remains an open problem.

As for approximation at the architecture level, we refer to Quora \cite{Venkataramani2013} as a representative example. Basically, the authors propose to extend the ISA of vector processors such that computation quality can be specified in the instruction set. In their proposal, error precision is deterministically bound for each instruction.

Finally, there is approximation in hardware circuits. There are several proposals in the literature for approximate arithmetic units like adders and multipliers. They can be generally classified to deterministic (e.g., adaptive reduced precision) and nondeterministic. Circuits of the latter type work most of the time as expected. However, they can occasionally produce arbitrary outputs.
The approximate circuits previously discussed are mostly designed manually for their specific purpose. In contrast, the authors of SALSA \cite{Venkataramani2012} approach hardware approximation from a different perspective. They propose a general technique to automatically synthesize approximate circuits given golden RTL model and quality specifications.

We conclude this section by discussing how cost is expected to increase as we go from algorithm level to circuit level. Given a (correct) algorithm that exposes some configurable knobs. Adapting such algorithm to different settings is relatively cheap. Also, it can be highly automated based on established meta-optimization literature as can be found in meta-learning.
However, things get more challenging if we were to approximate outputs based on the internal workings of an algorithm. 
Generally, this application level approximation requires asking users for annotations or assumptions on expected inputs.
Consequently, there is a smaller opportunity for automation and more difficulty in guaranteeing error quality.

The expected cost gets even higher as we move to architecture and circuit level approximation. Note that several stakeholders will be affected by such approximation. Compiler engineers, operating systems developers, and hardware architects all need to be either directly involved or at least aware of the approximation intended by the original algorithm developers. A proposed AC technique should demonstrate a serious value across the board to convince all of these people to get involved.

%%%%%%%%%%%%%%%%%%%%%%%%%%%%%%%%%%%%%%%%%%%%%%
\section{On nondeterministic approximate computing}
%%%%%%%%%%%%%%%%%%%%%%%%%%%%%%%%%%%%%%%%%%%%%%

Let's begin this section with a definition: An algorithm that may produce different outputs for the same input is considered to be nondeterministic. It's useful in our context to identify two categories of nondeterminism. In a \textit{partially} nondeterministic algorithm, nondeterminism sources can be feasibly accounted for a priori, i.e., they are a design choice with known consequences. Otherwise, the algorithm is considered to be \textit{fully} nondeterministic.
Simulated annealing is an example of a partially nondeterministic algorithm where choosing the next point in the search space is based on some randomness. Despite this, the control-flow behavior of the algorithm remains predictable. In contrast, in the case of fully nondeterministic algorithms, the followed control-flow path can differ in ``unexpected'' ways.

Several AC proposals have considered introducing some degree of nondeterminism to otherwise deterministic systems. In ApproxHadoop, the authors proposed random task dropping to gain more efficiency. Also, authors of SAGE \cite{Samadi2013}  proposed skipping atomic primitives to gain performance at the expense of exposing the algorithm to race conditions.
Nondeterministic AC introduced by unreliable hardware has received special attention from the research community. This is motivated by the potential efficiency gains discussed already. We shall focus on hardware-based nondeterministic AC in the following.

Nondeterministic AC can be implemented in several ways in hardware. DRAM cells need  periodic refresh to retain their data which consumes energy. Equipping DRAMs with ``selective'' no-refresh mechanisms saves energy but risks occasional bit errors.
Similarly, dynamically adapting bus compression and error detection mechanisms can provide significant gains in the communication between main memory and processing cores.

Additionally, there are efficiency opportunities in allowing the processor cores themselves to provide a best-effort rather than a reliable service. In this setting, hardware engineers may optimistically invest in reliability mechanisms.
However, software programmers need, in turn, to be aware of hardware unreliability and invest in fault management schemes suitable for their particular needs. Relax \cite{DeKruijf2010}  provide a good example of such an arrangement.
Also, CLEAR \cite{ChengMSCCSSLABM16} provides an interesting design-space exploration of reliability against soft-errors across the entire system stack.

Despite the extensive research in reliability in general and more recently in AC. The problem of running software \textit{reliably} and \textit{efficiently} on unreliable hardware is far from solved. Beside the cost mentioned in the previous section, there are still major interdisciplinary problems to be addressed.

First, there is the abstraction problem of the hardware/software interface. Extending the ISA abstraction makes sense given its ubiquity. For example, each ISA instruction might be extended with probability specification. This probability quantifies how many times this nondeterministic instruction is expected to supply correct results (this is a frequentist view, of course).
However, microchip designs nowadays are quite complex. They might comprise tens of IP modules from several IP providers. It is difficult for a microchip manufacturer to derive reliability probabilities per individual instruction. Also, even where such derivation is possible, microchip manufacturers would be reluctant to guarantee such probabilities to their customers.
Further, we believe that it is still not clear yet how much value can this instruction-level abstraction provide to hardware as well as software designs. Consequently, the question of suitable \textit{generic} abstractions between software and nondeterministic hardware is still open.

Second, there is the correctness problem. Can we establish that a software running on nondeterministic hardware is correct or perhaps probably correct? A short answer is probably no!
There are several reasons behind that: Statistical correctness requires sampling from (joint) probability distributions over inputs which are typically not available a priori. Also, software functions are generally noncontinuous and nonlinear in the mathematical sense. This makes them not only good at hiding unexpected behaviors in the corners, but also good at amplifying small error in unpredictable ways.
Additionally, algorithms process inputs in deterministic steps based on a specific control-flow path. Nondeterministic hardware might affect control-flow decisions causing the algorithm to immediately fail, or worse, proceed and produce arbitrary outputs.

We elaborate on the nondeterministic effects on control-flow based on the following code snippet. The function \texttt{approximate} takes two inputs \texttt{i1} and \texttt{i2} and one hyperparameter input \texttt{i3}. The latter is assumed to be a positive integer. 
From a reliability perspective, local parameters \texttt{k} and \texttt{i3} need to be protected (e.g., using software redundancy) otherwise the loop might not terminate.

\begin{lstlisting}[caption={Protecting \texttt{foo} and \texttt{bar} using software-based
reliability mechanisms is costly. However, we can't generally guarantee bounds on \texttt{result} without such mechanisms.},captionpos=b]
int approximate(int i1, int i2, int i3) {
  int result = 0;
  for(int k=0; k == i3; k++){
    if(foo(i1) < bar(i2))
      result += foo(i1)
    else
      result += bar(i2)
    }
    return result
}
\end{lstlisting}

The question now is what shall we do with functions \texttt{foo} and \texttt{bar}? Leaving them unprotected means that we risk producing unpredictable value in \texttt{result}. Note that a control-flow decision at line \#4 is based on a comparison between the results of \texttt{foo} and \texttt{bar}.
On the other hand, applying software-based reliability to \texttt{foo} and \texttt{bar} can prove more costly than running function \texttt{approximate} on deterministic hardware in the first place. This is the essence of control-flow wall. 
Basically, decisions taken in the control-flow usually depend on the processed data
which is costly to protect in software. 
On the other hand, allowing nondeterministic errors to affect the data means that we can't, in general, guarantee how the algorithm would behave.

In general, increasing reliability requires adding some redundancy to a given computing system. 
Such cost is considered acceptable in safety-critical systems where duplicating (or even tripling) hardware units is a common practice. 
Similarly, the cost of reliability is unavoidable for a distributed system to be fault tolerant. It's simply not feasible to keep thousands of computing nodes online all the time. We should plan for a fraction of the nodes to go offline, e.g., due to link failure, and remaining nodes should gracefully takeover their functionality. 
Nondeterministic hardware, in contrast to these settings, presents a different trade-off.
We should carefully reduce the cost of hardware reliability while incurring more
cost in terms of software reliability and overall system complexity.  
The latter cost should be maintained low enough for the approximate system to
operate more efficiently compared to the original system.

General-purpose programming on nondeterministic hardware was tackled in Chisel \cite{Misailovic2014} (A successor for a language called Rely). The authors assume a hardware model where processors provide reliable and unreliable versions of instructions. Also, data can be stored in an unreliable memory.
There, developers are expected to provide reliability specifications. Also, hardware engineers need to provide approximation specification. The authors combine static analysis and Integer-Linear Programming in order explore the design-space while maintaining the validity of reliability specification. Still, their analyses are limited by data dependencies in the control-flow graph.

It is important to differentiate between the reliability specification in Chisel and the similar probabilistic specification of say Uncertain<T> \cite{Bornholt2014}. The latter is a probabilistic programming extension to general-purpose languages. This means that the inputs are, typically, prior probability distributions that need to be processed \textit{deterministically}.

That said, and without being able to guarantee (probable) correctness, would it make any sense to use nondeterministic hardware? Well, it depends. In the case where the cost of failure is small, and errors can't propagate deep into the program anyway, being failure oblivious might make sense \cite{Rinard2004}.
Also, heterogeneous reliability can make nondeterministic hardware a more viable option in practice. Basically, reliable cores would be used to run control-dominated software, e.g., operating systems and language runtime. Only compute-intensive kernels can be offloaded to accelerators which are possibly built using unreliable nondeterministic hardware. A notable example here is ERSA \cite{Leem:2010a}.

%%%%%%%%%%%%%%%%%%%%%%%%%%%%%%%%%%%%%%%%%%%%%%
\section{Conclusion}
%%%%%%%%%%%%%%%%%%%%%%%%%%%%%%%%%%%%%%%%%%%%%%
Improving efficiency is a continuous endeavor in all engineering disciplines. This endeavor requires balancing trade-offs between system cost and gained value. The goal is to obtain results that are good enough for the cost we invest.
Approximate computing is the research area where we attempt to realize techniques to automatically gain computing efficiency by trading off output quality with a metric of interest such as performance and energy. Automation is key to the value proposal of approximate computing as practitioners are already capable of manually balancing those trade-offs.
In addition, approximation  needs to be principled which allow practitioners to trust the system to behave as expected in the real world. Combining automation and principled guarantees is essential, in our opinion, for approximate computing to have a secure place in the engineering toolbox.
This article briefly introduced approximate computing. The discussion covered the entire system stack from algorithms to hardware circuits. Also, we elaborated a bit on nondeterministic approximate computing due to the special attention it received from the research community.

\bibliographystyle{abbrvnat}
\bibliography{approx-intro-ref}

\clearpage
\end{document}